\documentclass[aoas,MSNbibl,nameyear,seceqn,dvips]{arximspdf}
\usepackage{multirow,dcolumn}
\usepackage{graphicx}

\doi{10.1214/13-AOAS684} 
\volume{8}
\issue{1}
\pubyear{2014}
\firstpage{74}
\lastpage{88}

\makeatletter
\newcommand{\rrvert}{\vert}
\newcommand{\llvert}{\vert}
\newcolumntype{d}[1]{D{.}{.}{#1}}
\newcommand{\var}{\operatorname{var}}

\makeatother

\begin{document}
\begin{frontmatter}

\title{Beta regression for time series analysis of bounded data,
with application to Canada Google\tsup{\textregistered}
Flu Trends}
\runtitle{Beta regression for time series analysis}
%

\begin{aug}
\author[a]{\fnms{Annamaria} \snm{Guolo}\ead
[label=e1]{annamaria.guolo@univr.it}}
\and
\author[a]{\fnms{Cristiano} \snm{Varin}\corref{}\ead[label=e2]{sammy@unive.it}}
\runauthor{A. Guolo and C. Varin}
\affiliation{Universit\`a di Verona and Universit\`a Ca' Foscari Venezia}
\address[a]{Department of Economics\\
Universit\`a di Verona\\
Via dell'Artigliere, 19\\
I-37129 Verona\\
Italy \\
\printead{e1}}

\address{Department of Environmental Sciences\\
\quad Informatics and Statistics\\
Universit\`a Ca' Foscari Venezia\\
San Giobbe Cannaregio, 873\\
I-30121 Venice\\
Italy \\
\printead{e2}}
\end{aug}

\received{\smonth{4} \syear{2013}}
\revised{\smonth{8} \syear{2013}}

%
\begin{abstract}
Bounded time series consisting of rates or proportions are often
encountered in applications. This manuscript proposes a practical
approach to analyze bounded time series, through a beta regression
model. The method allows the direct interpretation of the regression
parameters on the original response scale, while properly accounting
for the heteroskedasticity typical of bounded variables.
The serial dependence is modeled by a Gaussian copula, with a
correlation matrix corresponding to a stationary autoregressive and
moving average process. It is shown that inference, prediction, and
control can be carried out straightforwardly, with minor modifications
to standard analysis of autoregressive and moving average models.
The methodology is motivated by an application to the
influenza-like-illness incidence estimated by the
Google\tsup{\textregistered} Flu Trends project.
\end{abstract}

%
%
%
\end{frontmatter}

\section{Introduction}
Continuous bounded response variables, such as proportions and rates,
are frequently encountered in many areas of statistical practice. This
kind of data is usually examined through linear regression after a
logistic transformation. Despite its feasibility, such a modeling
strategy can suffer from some shortcomings, the most relevant being
that regression parameters are not directly interpretable on the
original response scale, as a consequence of Jensen's inequality. See
\citet{kieschnick03} and \citet{cribari10} for detailed discussions.

An alternative to linear modeling after logistic transformation
consists in a direct analysis of the bounded responses on their
original scale. To this purpose, the beta regression model has
attracted increasing interest in recent years, as a consequence of the
flexibility of the beta distribution in accommodating a variety of
distributional shapes over the unit interval.
Beta regression modeling of independent observations has been
illustrated in \citet{paolino01}, \citet{ferrari04}, and \citet
{smithson06}. Recent applications of beta regression in life sciences
have been encountered in clinical medicine [\citet{zou10,wang11}],
neuroscience [\citet{wang12}], pharmacometrics [\citet{rogers12}], and
virology [\citet{love10}].

Recent developments of beta regression analysis of bounded time series
have been addressed to observation-driven models [\citet
{rocha09,casarin12}] and to parameter-driven models [\citet{dasilva12}].
Straightforward likelihood inference makes the observation-driven model
appealing. A possible drawback arises in the case of regression
analysis, since the interpretation of the coefficients depends on past
transformed observations in the mean. Parameter-driven models are
attractive given their hierarchical construction. Nevertheless,
inference and prediction are complicated by the presence of correlated
latent variables.

As an alternative to the conditional observation- and parameter-driven
models, we suggest a marginal regression approach, through the
specification of a convenient class of beta regression models with
autoregressive and moving average errors. The serial dependence is
modeled by a Gaussian copula. Likelihood inference, prediction, and
control are carried out in a straightforward manner, with a
computational complexity similar to that of an ordinary ARMA model. In
addition, the approach allows an attractive interpretation of model components.

This article is motivated by surveillance of influenza through analysis
of the influenza-like-illness percentage estimated from aggregated web
search queries by the Google\tsup{\textregistered} Flu
Trends project. Analysis of influenza time series is a key step in
disease surveillance for monitoring the progress of epidemics, early
identification of pandemics, and ascertainment of factors associated to
unexpected changes in flu levels.

The plan of the article is as follows. Section~\ref{sec:google}
describes the motivating Google\tsup{\textregistered} Flu
Trends data. Section~\ref{sec:beta} summarizes beta regression modeling
and some extensions for time series analysis. The proposed methodology
is detailed in Section~\ref{sec:betamarg} and its finite sample
performance is investigated through simulation in Section~\ref{sec:simulations}. Section~\ref{sec:survei} describes online monitoring
of influenza outbreaks through control charts applied to beta
regression predictive quantile residuals. The application to the real
data set of interest is given in Section~\ref{sec:application}. Final
remarks in Section~\ref{sec:final} conclude.

Methods described in the paper are implemented within the more general
\texttt R [\citet{R12}] package \texttt{gcmr} ``Gaussian copula
marginal regression'' [\citet{masarotto12}], version 0.6.1. The package
is freely available at the CRAN repository, URL \href
{http://cran.r-project.org/web/packages/gcmr}{cran.r-project.org/web/packages/gcmr}.
Supplementary material [\citet{guolo13}] provides a brief illustration
of the \texttt R code.

\section{Motivating example}\label{sec:google}
The Google\tsup{\textregistered} Flu Trends project aims at
early detection of influenza-like-illness (ILI) activity around the
world. The ILI activity is measured in terms of cases per $100\mbox{,}000$
persons. The number of cases is reconstructed starting from aggregated
Google\tsup{\textregistered} search queries related to the
disease, such as, for example, \emph{influenza complication}, \emph{flu
remedy}, \emph{influenza symptoms}, and \emph{antiviral medication}.
See \citet{ginsberg09} for details about ILI counts estimation. The
Google\tsup{\textregistered} estimated ILI time series are
publicly available at URL \href{http://www.google.org/flutrends}{www.google.org/flutrends}. Data start on
the last week of 2002 for Brazil and Peru. Information has been
successively extended to 26 other countries all around the world.
Researchers at the U.S. Centers for Disease Control and Prevention
consider Google\tsup{\textregistered} Flu Trends as an early
warning of an outbreak, although not a substitute for traditional
epidemiological surveillance networks.
In fact, recent data from the U.S. indicate that peak influenza levels
in winter 2012--2013 have been overestimated, as a consequence of an
increased number of search queries related to influenza strains which
caused more serious illness and deaths than usual [\citet{butler13}].
%
\begin{figure}

\includegraphics{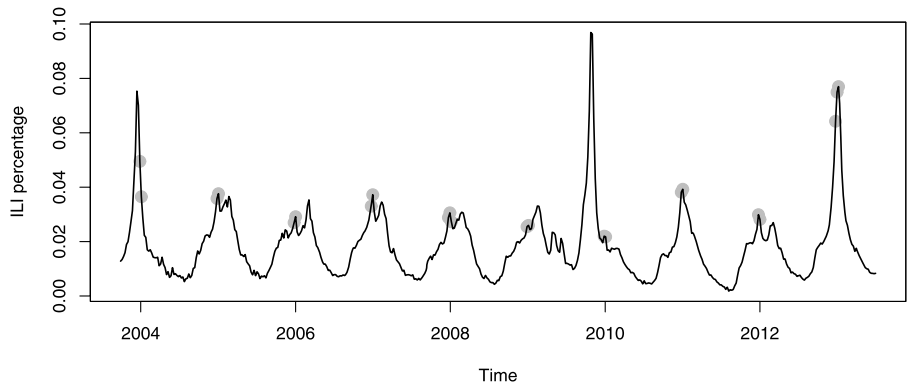}

\caption{Google\tsup{\textregistered} Flu Trends estimated
ILI percentage for Canada. Circles denote Christmas/New Year holidays.
Data source: \protect\href{http://www.google.org/flutrends}{www.google.org/flutrends}.}\label{fig:flu}
\end{figure}

Figure~\ref{fig:flu} displays the time series of
Google\tsup
{\textregistered} estimated ILI percentage, obtained as estimated ILI
counts divided by $100\mbox{,}000$ persons, for Canada. The time series
covers $510$ consecutive weeks in the period October 2003--June 2013.
Canada has been chosen since Google\tsup{\textregistered}
estimated ILI percentage highlights three epidemic peaks in December
2003, October--November 2009, and December 2012--January 2013. In
these periods, ILI peaked at about 7.5\%, 9.7\%, and 7.7\% of
Canadians, respectively, against normal seasonal influenza peaks of
about 3.5\%.

\section{Beta regression}\label{sec:beta}
Let $Y_t$ be a response variable bounded on the unit interval $(0,1)$,
$t=1, \ldots, n$, and let $\mathbf{x}_t$ be a vector of $p$ concomitant
covariates. According to \citet{paolino01} and \citet{ferrari04}, beta
regression assumes that $Y_t$ given $\mathbf{x}_t$ follows a beta
distribution $\operatorname{Beta}(\mu_t, \kappa_t)$ parametrized in terms of
the mean parameter $0< \mu_t < 1$ and the precision parameter $\kappa
_t>0$. It follows that $\operatorname{var}(Y_t)=\mu_t(1-\mu_t)/(1+\kappa_t)$
and the density function of $Y_t$ is
%
%
\begin{equation}
\label{eq:density} p_t(y_t; \bolds{\beta})=\frac{\Gamma(\kappa_t)}{\Gamma(\mu_t \kappa_t)
\Gamma \{ (1-\mu_t) \kappa_t \}}
y_t^{ \mu_t \kappa_t-1} (1-y_t)^{(1-\mu_t) \kappa_t-1},
\end{equation}
where $\Gamma(\cdot)$ denotes the Gamma function and subscript $t$ in
$p_t(\cdot)$ emphasizes the time dependence of the beta density through
$\mu_t$ and $\kappa_t$.

Dependence of the response $Y_t$ on the covariates $\mathbf{x}_t$ is
obtained by assuming a logit-linear model for the mean parameter,
$\operatorname{logit}(\mu_t)=\mathbf{x}_t^\top\bolds{\beta}_{\mathbf{x}}$, where
$\bolds
{\beta}_{\mathbf{x}}$ is a $p$-dimensional vector of coefficients.
Alternative link functions $g\dvtx (0,1) \rightarrow\mathbb{R}$ are
allowed, provided that they are monotonic and differentiable, such as,
for example, probit and log--log. Since the distribution of bounded
variables is characterized by heterogeneity, it is reasonable to model
the precision parameter with a log-linear model $\operatorname{log}(\kappa
_t)= \mathbf{z}_t^\top\bolds{\beta}_{\mathbf{z}}$, where $\mathbf{z}$ is a set
of $q$
covariates with associated vector of coefficients $\bolds{\beta}_{\mathbf
{z}}$. Implementations of beta regression analysis for independent
observations are available through \texttt{R} packages \texttt{betareg}
[\citet{cribari10,grun12}] and \texttt{gamlss} [\citet{stasino07}].

Within the time series framework, serial correlation in nonlinear
regression analysis can be accounted for through conditional or
marginal models. Following \citet{cox81}, conditional models are
further classified as observation- and parameter-driven models. \citet
{rocha09} consider observation-driven beta regression
models where the response $Y_t$ is modeled as a function of past information,
\[
Y_{t} |\{ y_{t-1}, \ldots, y_1 \} \sim
\operatorname{Beta}(\mu_t, \kappa_t),
\]
with $\mu_t$ depending on both covariates $\mathbf{x}_t$ and
logit-transformed past observations through the $\operatorname{ARMA}(p,q)$ model
\[
\operatorname{logit}(\mu_t)= \mathbf{x}_t^\top\bolds{
\beta}_{\mathbf{x}} + \sum_{i=1}^p
\psi_t \bigl\{ \operatorname{logit}(y_{t-i}) -
\mathbf{x}_{t-i}^\top \bolds {\beta}_{\mathbf{x}} \bigr\} + \sum
_{j=1}^q \lambda_j
\varepsilon_{t-j}.
\]
In the expression above, $\varepsilon_t$ is a random error and $\bolds{\psi
}=(\psi_1, \ldots, \psi_p)^\top$ and $\bolds{\lambda}=(\lambda_1,
\ldots,
\lambda_q)^\top$ are the autoregressive and moving average parameter
vectors, respectively. Straightforward likelihood inference makes the
observation-driven model appealing, although the interpretation of the
regression coefficients is complicated by the presence of past
transformed observations in the mean. \citet{casarin12} develop
Bayesian inference for purely autoregressive beta regression
observation-driven models and discuss selection of the optimal order.

\citet{dasilva12} investigate parameter-driven beta regression models,
extending \citet{dasilva11}. \citet{dasilva12} suppose responses
distributed as independent beta random variables conditionally on
latent variables.
Serial correlation is accounted for by assuming that the latent
variables evolve in time according to a state-space model. Although the
hierarchical model construction is attractive, likelihood computation
is complicated by the presence of $n$ correlated latent variables.
Likelihood approximation can be based on sequential simulation methods,
such as, for example, the Markov chain Monte Carlo approach discussed
by \citet{dasilva12}.

\section{Marginal beta regression time series modeling}\label{sec:betamarg}
In this paper we develop a marginal\vadjust{\goodbreak} extension of the beta regression
model for time series analysis which avoids the difficulties of
interpretation of observation-driven models and the computational
complications of parameter-driven models. Thereafter, the cumulative
distribution function of a normal variable with mean $m$ and variance
$s^2$ will be denoted by $\Phi(\cdot; m, s)$. A similar notation will
be used for the density function $\phi(\cdot; m, s)$. The common
simplified notation $\Phi(\cdot)=\Phi(\cdot; 0, 1)$ and $\phi
(\cdot
)=\phi(\cdot; 0, 1)$ is adopted for a standard normal variable.

The proposed marginal beta regression model exploits the probability
integral transformation to relate response $Y_t$ to covariates $\mathbf
{x}_t$ and $\mathbf{z}_t$ and to a standard normal error $\varepsilon_t$,
%
%
\begin{equation}
\label{eq:mod1} Y_t=F_t^{-1} \bigl\{ \Phi(
\varepsilon_t); \bolds{\beta} \bigr\},
\end{equation}
where $F_t(\cdot; \bolds{\beta})$ is the cumulative distribution function
associated to density~(\ref{eq:density}), $\bolds{\beta}=(\bolds{\beta
}_{\mathbf
{x}}^\top, \bolds{\beta}_{\mathbf{z}}^\top)^\top$. The probability integral
transformation implies that $Y_t$ is \emph{mar\-ginally} beta
distributed, $Y_t\sim\operatorname{Beta}(\mu_t, \kappa_t)$. Remaining serial
correlation not accounted for by covariates $\mathbf{x}_t$ and $\mathbf{z}_t$
is modeled by assuming that errors $\varepsilon_t$ follow a stationary
$\operatorname{ARMA}(p,q)$ process,
%
%
\begin{equation}
\label{eq:mod2} \varepsilon_t=\sum_{i=1}^p
\psi_i \varepsilon_{t-i}+\sum_{j=1}^q
\lambda_j \eta_{t-j} +\eta_t,
\end{equation}
where $\eta_t$ are independent zero-mean normal variables. In order to
assure $\varepsilon_t$ having unit variance, the variance of $\eta_t$ is
an appropriate function of the autoregressive parameter vector $\bolds
{\psi
}$ and the moving average parameter vector $\bolds{\lambda}$. For example,
if errors follow the AR$(1)$ process
$\varepsilon_t=\psi\varepsilon_{t-1}+\eta_t$,
then $\var(\eta_t)=1-\psi^2$.

The proposed beta regression model expressed by equations (\ref{eq:mod1})--(\ref{eq:mod2}) has the advantage of separating the time
series component $\varepsilon_t$ from the regression part. This allows a
straightforward interpretation of the regression coefficients as if
observations were independent. Models (\ref{eq:mod1})--(\ref{eq:mod2}) is
an instance of Gaussian copula marginal regression [\citet{song07},
Chapter~6; \citet{masarotto12}].

Let $\bolds{\theta}$ denote the whole parameter vector formed by the
regression parameter vector $\bolds{\beta}$ and the ARMA parameter vectors
$\bolds{\psi}$ and $\bolds{\lambda}$. Inference on $\bolds{\theta}$, diagnostics
of departures from model assumptions, and prediction of future outcomes
require the specification of the $k$-lags ahead predictive density
$p_{t+k}(y_{t+k}| y_t, \ldots, y_1; \bolds{\theta})$. Such a density can
be obtained by standard transformation rules as the product of the
$k$-lags ahead predictive density of the errors and the Jacobian of the
transformation $\varepsilon_{t+k}=\Phi^{-1}\{ F_{t+k}(y_{t+k}; \bolds
{\beta
})\}$,
%
%
\begin{eqnarray}
\label{eq:predictive}
\nonumber
p_{t+k}(y_{t+k}| y_t,
\ldots, y_1; \bolds{\theta})&=& p(\varepsilon_{t+k}|
\varepsilon_t, \ldots, \varepsilon_1;\bolds{\theta}) \biggl
\llvert \frac{d
\varepsilon
_{t+k}}{d y_{t+k}} \biggr\rrvert
\nonumber\\
&=& p_{t+k}(y_{t+k}; \bolds{\beta}) \frac{p(\varepsilon_{t+k}| \varepsilon_t,
\ldots, \varepsilon_1; \bolds{\theta}) }{p(\varepsilon_{t+k}; \bolds{\beta
})}
\\
&=& p_{t+k}(y_{t+k}; \bolds{\beta}) \frac{\phi(\varepsilon_{t+k};
m_{t+k|t}, s_{t+k|t})}{\phi(\varepsilon_{t+k})},\nonumber
\end{eqnarray}
where $m_{t+k|t}=\mathrm{E}(\varepsilon_{t+k}| \varepsilon_t, \ldots,
\varepsilon
_1; \bolds{\theta})$ and $s^2_{t+k|t}=\operatorname{var}(\varepsilon_{t+k}|
\varepsilon
_t, \ldots, \varepsilon_1; \bolds{\theta})$. Both conditional expectations
can be efficiently evaluated in a linear number of operations via
Kalman filter recursions.

\begin{figure}

\includegraphics{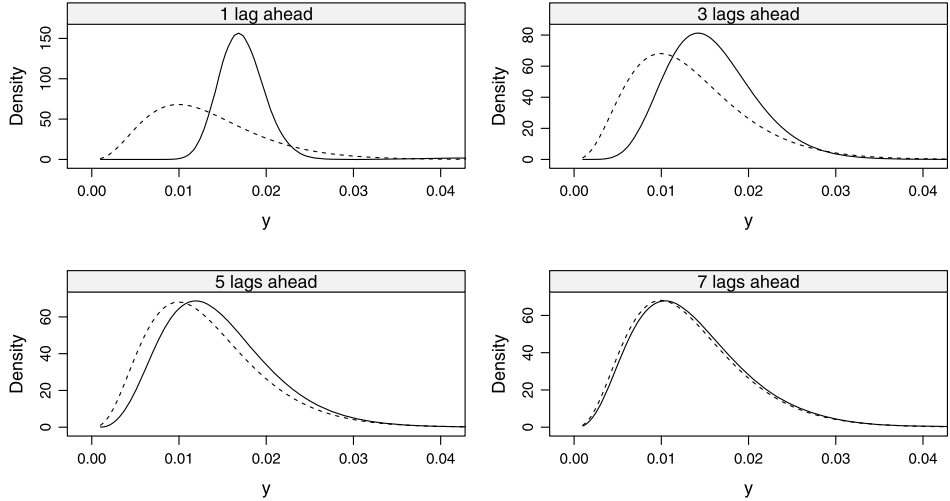}

\caption{Predictive density (solid line) and marginal density (dashed
line) at different lags ahead for the marginal beta regression model
with $\operatorname{ARMA}(2,1)$ errors described in the simulation study,
Section~\protect\ref{sec:simulations}.}\label{fig:pred_simul}
\end{figure}

Expression (\ref{eq:predictive}) is particularly attractive in terms of
interpretability, since it separates the marginal density associated to
the future observation, $p_{t+k}(y_{t+k}; \bolds{\beta})$, from a measure
of the serial correlation within the errors. Figure~\ref{fig:pred_simul} provides an illustration of the beta regression model
with $\operatorname{ARMA}(2,1)$ errors used for the simulation study in Section~\ref{sec:simulations}. The marginal density $p_{t+k}(y_{t+k}; \bolds{\beta})$
and the predictive density $p_{t+k}(y_{t+k}| y_t, \ldots, y_1; \bolds
{\theta})$ substantially differ for short time prediction, with the
predictive density being more peaked since it accounts for the
information in the past observations. As the prediction lag increases,
past data become less informative, thus making the predictive density
closer to the marginal density, as expected.
%

Basic properties of the $\operatorname{ARMA}(p,q)$ process are inherited by the
proposed model. In fact, it is immediate from (\ref{eq:predictive})
that if errors $\varepsilon_t$ follow a $\operatorname{MA}(q)$ process, then
observations more than $q$ units far apart are independent. Moreover,
if errors $\varepsilon_t$ follow an $\operatorname{AR}(p)$ process, then
observations follow a Markovian process of order $p$.

By model construction, the predictive cumulative distribution function
of $Y_{t+k}$ given $\{y_t, \ldots, y_1\}$ coincides with the predictive
cumulative\ distribution function of $\varepsilon_{t+k}$ given $\{\varepsilon
_t, \ldots, \varepsilon_1\}$,
%
%
\begin{eqnarray}
\label{eq:cdf_pred}
\nonumber
F_{t+k}(y_{t+k}| y_t,
\ldots, y_1; \bolds{\theta})&=& \int_{0}^{y_{t+k}}
p_{t+k}(u| y_t, \ldots, y_1; \bolds{\theta}) \,d u
\nonumber\\
&=&\int_{-\infty}^{\Phi^{-1}\{F_{t+k}(y_{t+k}; \bolds{\beta})\}} p(\varepsilon
_{t+k}| \varepsilon_t, \ldots, \varepsilon_1;
\bolds{\theta}) \,d \varepsilon _{t+k}
\\
&=&\Phi ( \varepsilon_{t+k}; m_{t+k|t}, s_{t+k|t} ).\nonumber
\end{eqnarray}
Accordingly, the $\alpha$-quantile of the predictive distribution is
\[
y_{t+k| t; \alpha}=F_{t+k}^{-1} \bigl[\Phi \bigl
\{m_{t+k| t}+\Phi ^{-1}(\alpha) s_{t+k|t} \bigr\}; \bolds{
\beta} \bigr].
\]

\subsection{Likelihood inference}\label{sec:lik}
We suggest to perform inference by relying on maximum likelihood
estimation. Let ${L}_{\mathrm{ind}}(\bolds{\beta}; \mathbf{y})=\prod_{t=1}^n
p_t(y_t; \bolds{\beta})$ denote the likelihood constructed under the
assumption of independence. Then, given the result in (\ref{eq:predictive}), the likelihood function for $\bolds{\theta}$ is
\begin{eqnarray*}
\label{eq:like}
\nonumber
{L}(\bolds{\theta}; \mathbf{y})&=& p_1(y_1;
\bolds{\beta}) \prod_{t=2}^{n}
p_{t}(y_{t}| y_{t-1}, \ldots, y_1;
\bolds{\theta})
\\
&=& {L}_{\mathrm{ind}}(\bolds{\beta}; \mathbf{y}) \prod
_{t=2}^n \frac
{p(\varepsilon_t | \varepsilon_{t-1}, \ldots, \varepsilon_1; \bolds{\theta
})}{p(\varepsilon_t; \bolds{\beta})}.
\end{eqnarray*}
The likelihood function is the product of the independence likelihood $
{L}_{\mathrm{ind}}$ and a calibration term accounting for the presence
of dependence of $\varepsilon_t$ on past values. A~calibration term
significantly different from one is indicative of dependence.

From a practical point of view, the closed-form of the likelihood
implies an effortless computation. As already noted for the predictive
density, the Kalman filter can be employed for efficient computation of
the predictive densities of the $\operatorname{ARMA}(p,q)$ errors, $p(\varepsilon
_{t} | \varepsilon_{t-1}, \ldots, \varepsilon_1; \bolds{\theta})$, thus making
the computational complexity of likelihood evaluation of a linear order.

\subsection{Predictive quantile residuals}\label{sec:diagnostics}
Following \citet{dunn96} and \citet{masarotto12}, model validation can
be based on the analysis of the predictive quantile residuals
\[
r_t=\Phi^{-1} \bigl\{F_{t}(y_{t}|y_{t-1},
\ldots, y_1; \hat{\bolds {\theta }}) \bigr\},
\]
where $\hat{\bolds{\theta}}$ denotes the maximum likelihood estimate of
$\bolds{\theta}$.
Given (\ref{eq:cdf_pred}), predictive quantile residuals $r_t$ assume
the familiar form
\[
r_t=\frac{\hat\varepsilon_t-\hat m_{t| t-1}}{\hat s_{t|t-1}},
\]
where $\hat\varepsilon_t$, $\hat m_{t|t-1}$, and $\hat s_{t|t-1}$ are
evaluated at $\hat{\bolds{\theta}}$. Residuals $r_t$ are realizations of
$n$ independent standard normal variables if the model assumptions are met.

\section{Simulation study}\label{sec:simulations}
A simulation study has been performed in order to evaluate maximum
likelihood estimation and prediction for the proposed marginal beta
regression model. The simulation setup consists of 1000 weekly time
series from the marginal beta regression model specified as follows.
The length of the time series is set equal to 368, with the first
$n=52\times7=364$ observations used for model fitting and the
remaining four observations used for prediction. Following common
practice in surveillance literature [\citet{unkel12}], mean $\mu_t$ and
precision $\kappa_t$ include linear trend and annual seasonal
components representing temperature variations,
%
%
\begin{eqnarray}
\label{eq:model}
\nonumber
\operatorname{logit}(\mu_t)&=&
\beta_{0\mathbf{x}}+\beta_{1\mathbf{x}} \tilde {t}+ \beta _{2\mathbf{x}}\sin
\biggl(\frac{2\pi t}{52} \biggr)+ \beta_{3\mathbf
{x}}\cos \biggl(
\frac{2\pi t}{52} \biggr),
\nonumber
\\[-8pt]
\\[-8pt]
\nonumber
\log(\kappa_t)&=&\beta_{0\mathbf{z}}+\beta_{1\mathbf{z}}
\tilde{t}+ \beta_{2\mathbf
{z}}\sin \biggl(\frac{2\pi t}{52} \biggr)+
\beta_{3\mathbf{z}}\cos \biggl(\frac
{2\pi t}{52} \biggr),
\end{eqnarray}
where $\tilde t$ indicates the time index $t$ centered and scaled by
factor $100$ in such a way to avoid numerical instabilities.
The residual serial correlation is modeled by assuming an $\operatorname{ARMA}(2,1)$
process for the errors. The values of the parameters are set equal to
$\beta_{0\mathbf{x}}=-4.00$, $\beta_{1\mathbf{x}}=0.15$, $\beta_{2\mathbf
{x}}=-0.22$, $\beta_{3\mathbf{x}}=-0.67$, $ \beta_{0\mathbf{z}}=6.00$,
$\beta
_{1\mathbf{z}}=0.10$, $\beta_{2\mathbf{z}}=-0.06$, $\beta_{3\mathbf{z}}=-0.19$,
$\psi_1=1.50$, $\psi_2=-0.60$, and $\lambda=-0.30$. The values of
$\beta
_{2\mathbf{x}}$, $\beta_{3\mathbf{x}}$, $\beta_{2\mathbf{z}}$, and $\beta
_{3\mathbf
{z}}$ are chosen in order to guarantee an amplitude equal to $0.7$ and
$0.2$ for the mean and the precision, respectively, and a phase shift
equal to $0.6\pi$ for both mean and precision. These values resemble a
typical ILI weekly time series.

Table~\ref{tab:simul} displays average and standard deviation of the
parameter estimates, and average of the standard errors computed from
the inverse of the observed Fisher information. The results are
satisfactory, as they show (i) a negligible bias in the estimation of
all the parameters and (ii) averages of the standard errors close to
standard deviations of the estimates.
%
\begin{table}
\caption{Average (\texttt{ave}), standard deviation (\texttt{s.d.}),
and average of standard errors (\texttt{s.e.}) for 1000 simulated
estimates based on a beta regression model with $\operatorname{ARMA}(2,1)$ errors and
with independent errors}\label{tab:simul}
\begin{tabular*}{\textwidth}{@{\extracolsep{\fill}}lcd{2.2}d{2.2}d{1.2}d{1.2}d{2.2}d{1.2}d{1.2}@{}}
\hline
&& & \multicolumn{3}{c}{$\bolds{\operatorname{ARMA}(2,1)}$} &\multicolumn{3}{c}{\textbf{Independence}}
\\[-6pt]
&& & \multicolumn{3}{c}{\hrulefill} &\multicolumn{3}{c@{}}{\hrulefill}
\\
&& \multicolumn{1}{c}{\textbf{true}} & \multicolumn{1}{c}{\textbf{ave}} & \multicolumn{1}{c}{\textbf{s.d.}} & \multicolumn{1}{c}{\textbf{s.e.}} &
\multicolumn{1}{c}{\textbf{ave}} & \multicolumn{1}{c}{\textbf{s.d.}} & \multicolumn{1}{c@{}}{\textbf{s.e.}}\\
\hline
{Mean} &intercept & -4.00 & -4.01 & 0.06 & 0.05 &
-4.01 & 0.06 & 0.02 \\
&trend & 0.15 & 0.15 & 0.05 & 0.04 & 0.15 & 0.05 & 0.02 \\
&cosine term & -0.22 & -0.22 & 0.07 & 0.06 & -0.22 & 0.07 & 0.02 \\
&sine term & -0.67 & -0.67 & 0.08 & 0.07 & -0.67 & 0.08 & 0.03 \\[3pt]
{Precision} & intercept & 6.00 & 6.11 & 0.17 & 0.17 &
6.15 & 0.18 & 0.08 \\
& trend & 0.10 & 0.10 & 0.07 & 0.07 & 0.12 & 0.18 & 0.07 \\
&cosine term & -0.06 & -0.06 & 0.11 & 0.11 & -0.06 & 0.24 & 0.10 \\
&sine term& -0.19 & -0.20 & 0.11 & 0.11 & -0.22 & 0.25 & 0.11 \\[3pt]
{Errors} &ar1 & 1.50 & 1.51 & 0.12 & 0.11 & \multicolumn{1}{c}{--} & \multicolumn{1}{c}{--} &
\multicolumn{1}{c}{--} \\
&ar2 & -0.60 & -0.62 & 0.11 & 0.09 & \multicolumn{1}{c}{--} & \multicolumn{1}{c}{--} & \multicolumn{1}{c}{--} \\
&ma1 & -0.30 & -0.33 & 0.15 & 0.13 & \multicolumn{1}{c}{--} & \multicolumn{1}{c}{--} & \multicolumn{1}{c}{--}\\
\hline
\end{tabular*}
\end{table}

Table~\ref{tab:pred_simul} reports the empirical coverage of
prediction intervals at lags one to four, either for the fitted model
with $\operatorname{ARMA}(2,1)$ errors or for the independence model. Prediction
intervals from the model with $\operatorname{ARMA}(2,1)$ errors are sensibly closer to
the nominal level than those based on the independence model.
%
\begin{table}[b]
\caption{Empirical coverage of prediction intervals at various lags
ahead for 1000 simulated time series based on a beta regression model
with $\operatorname{ARMA}(2,1)$ errors and with independent errors}\label{tab:pred_simul}
\begin{tabular*}{\textwidth}{@{\extracolsep{\fill}}lccccccccc@{}}
\hline
& & \multicolumn{4}{c}{$\bolds{\operatorname{ARMA}(2,1)}$} &\multicolumn{4}{c@{}}{\textbf{Independence}}
\\[-6pt]
& & \multicolumn{4}{c}{\hrulefill} &\multicolumn{4}{c@{}}{\hrulefill}
\\
& & \textbf{lag 1} & \textbf{lag 2} & \textbf{lag 3}& \textbf{lag 4} & \textbf{lag 1} & \textbf{lag 2} & \textbf{lag 3} & \textbf{lag 4} \\
\hline
{Levels} & 90\% & 0.895 & 0.886 & 0.870 & 0.885 & 0.880
& 0.868 & 0.857 & 0.851 \\
& 95\% & 0.948 & 0.933 & 0.930 & 0.930 & 0.932 & 0.932 & 0.913 & 0.900
\\
& 99\% & 0.985 & 0.985 & 0.978 & 0.973 & 0.971 & 0.970 & 0.956 & 0.948
\\
\hline
\end{tabular*}
\end{table}

\section{Monitoring outbreaks of disease}\label{sec:survei}
Quality control charts are typically employed for online detection of
outbreaks of infectious diseases, {for example}, \citet
{woodall06} and \citet{unkel12}. To this aim, the first step is the
identification of a model describing the pattern of ordinary influenza
seasons. Then, departures from the model-expected influenza levels are
interpreted as symptoms of anomalies. Cumulative sum (CUSUM) charts
[\citet{montgomery09}, Chapter~9] are appropriate for monitoring
long-lasting illnesses such as ILI, given the capability of early
detection of small variations in the mean disease level. In fact, CUSUM
charts are employed by the Centers for Disease Control and Prevention
for routinely syndromic surveillance [\citet{hutwagner03}].

CUSUM charts are typically constructed under the assumption of
independent observations from a normal distribution, at least
approximately. Accordingly, below we suggest to monitor influenza
disease through predictive quantile residuals~$r_t$.
The bilateral CUSUM chart is based on the positive $C_t^+$ and the
negative $C_t^-$ cumulative sums of $r_t$,
\begin{eqnarray*}
C_t^+&=&\max\bigl\{0, r_t-k+C_{t-1}^+ \bigr\},
\\
C_t^-&=&\max\bigl\{0, -k-r_t+C_{t-1}^- \bigr\}
\end{eqnarray*}
for a \emph{reference value} $k$ and with $C_0=0$. The process is
out-of-control if either $C_t^+$ or $C_t^-$ exceeds the \emph{decision
limit} $h$. Parameters $k$ and $h$ are chosen in order to guarantee an
acceptable capability to detect influenza levels anomalies and, in the
meanwhile, a low number of false alarms. Following standard
recommendations in quality control literature [\citet{montgomery09}],
the chart parameters can be set to values $k=0.5$ and $h=4$.

Standard application of CUSUM charts involves two phases. In Phase I,
historical data are analyzed to calibrate the chart
when the process is under control. Phase II is the online monitoring
stage based on the chart calibrated at the previous phase. Details are
given below:
\begin{enumerate}
\item Phase I
\begin{enumerate}[(a)]
\item[(a)] Fit the beta marginal regression model including trend,
seasonality, and $\operatorname{ARMA}(p,q)$ errors, with $p$ and $q$ large enough to
guarantee residual autocorrelation to be captured. As a rule of thumb,
we suggest $p=q=3$.
\item[(b)] Remove the anomalous observations identified by a CUSUM chart of
the predictive quantile residuals derived from the model fitted at step (a).
\item[(c)] Re-estimate the beta marginal regression model on the time series
without the anomalous observations. Choose the most appropriate
$\operatorname{ARMA}(p,q)$ structure, $p\leq3$ and $q\leq3$, via information
criteria or cross-validation. The chosen model is the best model
representation of a regular seasonal influenza.
\end{enumerate}
\item Phase II
\begin{enumerate}[(d)]
\item[(d)] Online monitor influenza outbreaks by the unilateral
positive CUSUM chart of the predictive quantile residuals derived from
the model selected at Phase I, step (c).
\end{enumerate}
\end{enumerate}

\section{Application to Canada Google\tsup{\textregistered}
Flu Trends}\label{sec:application}
In this section we illustrate the application of the methodology
previously described to the analysis of Canada Google\tsup
{\textregistered} Flu Trends data.

In order to illustrate the surveillance procedure of Section~\ref{sec:survei}, we used data until June 2010 for model calibration (Phase
I), while the following three years of observations are used for online
monitoring (Phase II). The initial CUSUM chart based on the $\operatorname{ARMA}(3,3)$
model in Phase I identifies 19 anomalous observations over 354
observations. The subsequent step is the estimation of all possible
models with $\operatorname{ARMA}(p,q)$ errors, $p \leq3$ and $q \leq3$, to the data
after removal of the 19 anomalous observations.
Table~\ref{tab:canada_aic} ranks the sixteen possible models in terms
of Akaike Information Criterion. The preferred model is the one with
$\operatorname{ARMA}(2,1)$ errors. However, results highlight that a precise
identification of $p$ and $q$ is not crucial, since many models induce
essentially the same autocorrelation structure; see Table~\ref{tab:canada_aic}.
%
\begin{table}
\caption{Canada Google\tsup{\textregistered} Flu Trends
data. Estimated beta marginal regression models with $\operatorname{ARMA}(p,q)$ errors
ranked according to the Akaike Information Criterion (AIC) and
corresponding autocorrelation of the errors at lags one to four}\label{tab:canada_aic}
\begin{tabular*}{\textwidth}{@{\extracolsep{\fill}}lccd{5.2}cccc@{}}
\hline
& \multicolumn{2}{c}{\textbf{ARMA}} & &
\multicolumn{4}{c@{}}{\textbf{Autocorrelations}}\\[-6pt]
& \multicolumn{2}{c}{\hrulefill} & & \multicolumn{4}{c@{}}{\hrulefill}\\
\textbf{Rank} & \multicolumn{1}{c}{$\bolds{{p}}$} & \multicolumn{1}{c}{$\bolds{{q}}$} & \multicolumn{1}{c}{\textbf{AIC}} & \textbf{lag 1} & \textbf{lag 2} & \textbf{lag 3} & \textbf{lag 4} \\
\hline
\phantom{0}1 & 2 & 1 & -3372.45 & 0.94 & 0.84 & 0.74 & 0.64 \\
\phantom{0}2 & 3 & 0 & -3372.37 & 0.94 & 0.84 & 0.74 & 0.64 \\
\phantom{0}3 & 2 & 0 & -3371.57 & 0.94 & 0.84 & 0.75 & 0.66 \\
\phantom{0}4 & 1 & 2 & -3371.47 & 0.94 & 0.84 & 0.74 & 0.66 \\
\phantom{0}5 & 3 & 1 & -3370.49 & 0.94 & 0.84 & 0.74 & 0.64 \\
\phantom{0}6 & 2 & 2 & -3370.46 & 0.94 & 0.84 & 0.74 & 0.64 \\
\phantom{0}7 & 1 & 3 & -3369.77 & 0.94 & 0.84 & 0.74 & 0.65 \\
\phantom{0}8 & 3 & 2 & -3368.66 & 0.94 & 0.84 & 0.74 & 0.64 \\
\phantom{0}9 & 2 & 3 & -3367.87 & 0.94 & 0.84 & 0.74 & 0.65 \\
10 & 3 & 3 & -3367.23 & 0.94 & 0.84 & 0.74 & 0.64 \\
11 & 1 & 1 & -3366.89 & 0.93 & 0.85 & 0.77 & 0.70 \\
12 & 1 & 0 & -3353.23 & 0.93 & 0.87 & 0.81 & 0.75 \\
13 & 0 & 3 & -3269.01 & 0.78 & 0.42 & 0.12 & 0.00 \\
14 & 0 & 2 & -3185.59 & 0.68 & 0.24 & 0.00 & 0.00 \\
15 & 0 & 1 & -3038.51 & 0.49 & 0.00 & 0.00 & 0.00 \\
16 & 0 & 0 & -2766.91 & 0.00 & 0.00 & 0.00 & 0.00 \\
\hline
\end{tabular*}
\end{table}

\begin{figure}

\includegraphics{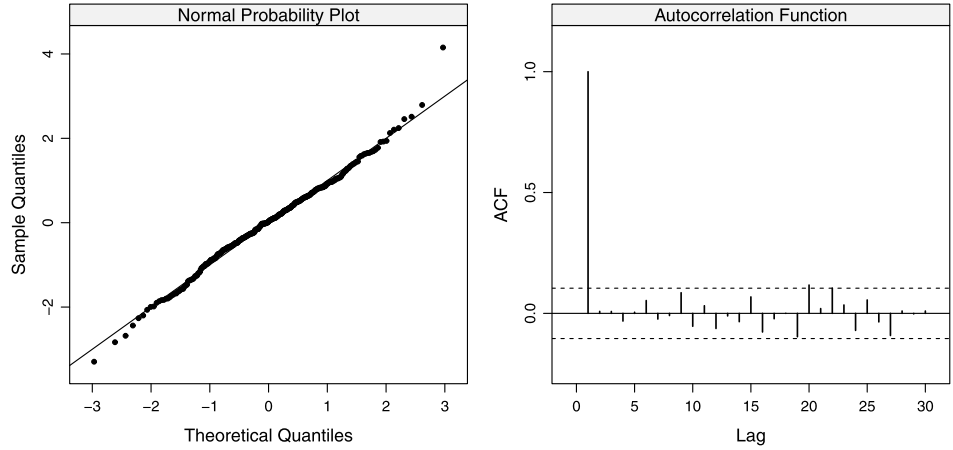}

\caption{Canada Google\tsup{\textregistered} Flu Trends
data. Normal probability plot (left panel) and autocorrelation function
(right panel) of the predictive quantile residuals for the fitted
marginal beta regression model with $\operatorname{ARMA}(2,1)$ errors.}\label{fig:canada_resid}
\end{figure}

The application of the CUSUM chart in Phase II requires the predictive
quantile residuals being comparable to a set of independent normal
variables. The graphical examination of the predictive quantile
residuals reported in Figure~\ref{fig:canada_resid} sustains such a
requirement.
%

\begin{figure}

\includegraphics{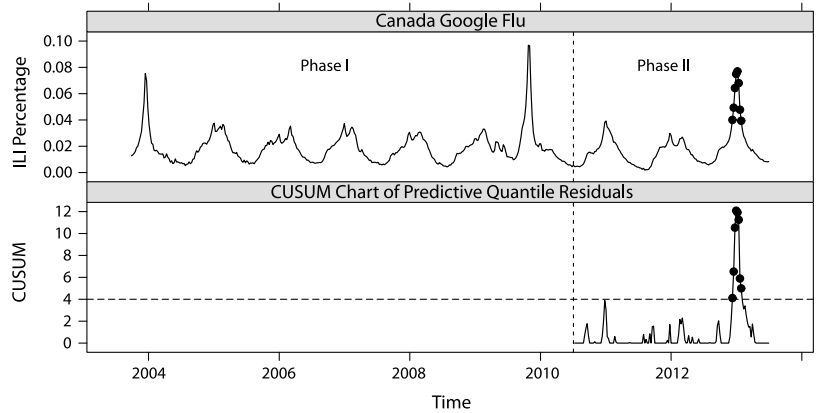}

\caption{Canada Google\tsup{\textregistered} Flu Trends
data. Positive CUSUM chart for surveillance of influenza outbreaks.
Circles indicate out-of-control weeks.}\label{fig:canada_cusum}
\end{figure}

Phase II CUSUM chart for online monitoring is illustrated in
Figure~\ref{fig:canada_cusum}. The corresponding points above the decision limit
$h=4$ in the influenza time series are highlighted in the bottom panel
of Figure~\ref{fig:canada_cusum}. The process is under control until
December 9, 2012, and then it remains out-of-control for eight
consecutive weeks before returning under control. The out-of-control
weeks correspond to the epidemic peak that occurred in December 2012--January 2013.
%

\subsection{Holiday peaks}
As observed by a referee, Canada Google\tsup{\textregistered} Flu Trends data show a
peak--valley--peak pattern within a couple of
weeks at the beginning of most of the observed years; see Figure~\ref{fig:flu}. Accordingly, we investigated the presence of a ``holiday
effect,'' related to the Christmas/New Year period. Table~\ref{tab:canada_best} reports estimates and standard errors for the
parameters of the beta marginal regression model with trend, sine, and
cosine terms describing seasonal temperature variations, $\operatorname{ARMA}(2,1)$
errors, and the dummy variable for the holiday weeks.
Results indicate no significant trend in the mean, which is instead
significant for the precision. The annual seasonal component is highly
significant in both mean and precision, as expected. The analysis
confirms a very significant increase of ILI in correspondence with the
holiday weeks, given an estimated holiday effect parameter in the mean
equal to 0.11, with a standard error of 0.02. Conversely, there is no
significant effect in terms of precision (estimate 0.12, standard error 0.09).
%
\begin{table}
\caption{Canada Google\tsup{\textregistered} Flu Trends
data. Estimates and standard errors for the parameters of fitted
marginal beta regression model without and with holiday effect. Akaike
Information Criterion (AIC) statistic also reported}\label{tab:canada_best}
\begin{tabular*}{\textwidth}{@{\extracolsep{\fill}}lcd{2.2}cd{2.2}c@{}}
\hline
& & \multicolumn{2}{c}{\textbf{No holiday effect}} & \multicolumn{2}{c@{}}{\textbf{Holiday
effect}}\\[-6pt]
& & \multicolumn{2}{c}{\hrulefill} & \multicolumn{2}{c@{}}{\hrulefill}\\
& \textbf{Parameter} & \multicolumn{1}{c}{\textbf{est.}} & \textbf{s.e.} & \multicolumn{1}{c}{\textbf{est.}} & \textbf{s.e.} \\
\hline
{Mean} &
intercept & -4.14 & 0.05 & -4.14 & 0.05 \\
& trend & -0.16 & 0.33 & 0.05 & 0.33 \\
& sine term & 0.66 & 0.06 & 0.65 & 0.06 \\
& cosine term & -0.31 & 0.06 & -0.31 & 0.06 \\
& Christmas/New Year & \multicolumn{1}{c}{--} & \multicolumn{1}{c}{--} & 0.11 & 0.02
\\[3pt]
{Precision} &
intercept & 6.23 & 0.11 & 6.19 & 0.11 \\
& trend & 1.46 & 0.43 & 1.68 & 0.43 \\
& sine term & -0.48 & 0.09 & -0.37 & 0.10 \\
& cosine term & -0.04 & 0.10 & -0.08 & 0.09 \\
& Christmas/New Year & \multicolumn{1}{c}{--} & \multicolumn{1}{c}{--} & 0.12 & 0.09
\\[3pt]
{ARMA} &
ar1 & 1.52 & 0.07 & 1.57 & 0.06 \\
& ar2 & -0.60 & 0.07 & -0.64 & 0.06 \\
& ma1 & -0.25 & 0.09 & -0.28 & 0.08 \\[6pt]
&\multicolumn{1}{c}{AIC} & \multicolumn{2}{c}{$-5028.74$} & \multicolumn{2}{c}{$-5057.31$}
\\
\hline
\end{tabular*}
\end{table}

\begin{figure}[b]

\includegraphics{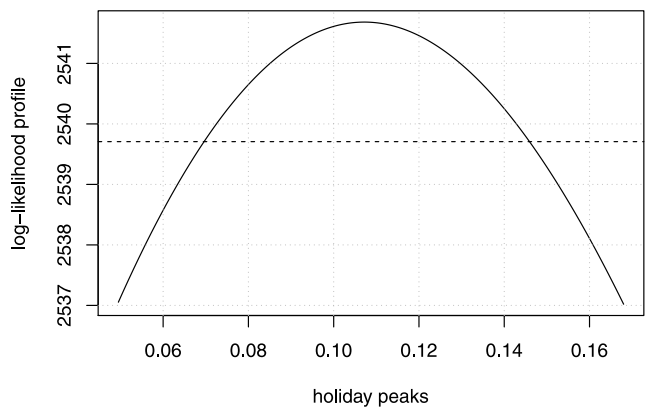}

\caption{Canada Google\tsup{\textregistered} Flu Trends
data. Profile log-likelihood for holiday effect parameter. Horizontal
dashed line corresponds to 95\% asymptotic confidence interval.}
\label{fig:canadaprofile}
\end{figure}

Further confirmations of the relevance of the holiday effect are
provided by AIC, which increases from $-5057.31$ to $-5028.74$, and by the
profile log-likelihood for the associated coefficient, displayed in
Figure~\ref{fig:canadaprofile}.
%

A brief illustration of how to use package \texttt{gcmr} for
replicating the analysis in this section is provided in the supplement
[\citet{guolo13}].

\section{Conclusions}\label{sec:final}
This paper suggested a practical approach for analysis of bounded time
series defined on the unit interval. One of the advantages of the
proposed marginal model is the reproducible interpretation of the
regression parameters, whose meaning does not depend on the ARMA
structure. The robust interpretation of the regression parameters is a
property not shared by alternative conditionally specified models, such
as observation- and parameter-driven beta regression models briefly
described in Section~\ref{sec:beta}. Another advantage of the proposed
approach is that inferential and prediction tasks have convenient
expressions, thus making modeling time series on the unit scale
feasible as a practical alternative to the common logit-transformation approach.

Several extensions of the proposed modeling framework are possible.
First, the approach has a trivial extension to time series defined on
an arbitrary $(a,b)$ interval. Second, spatial and spatio-temporal beta
regression models can be constructed by assuming that the errors are
realizations of a Gaussian random field. Finally, the model can be
extended to allow for exact zeros and ones, by using the zero-or-one
beta inflated regression model [\citet{ospina12}] to define the
univariate marginal distributions.

\section*{Acknowledgments}
The authors desire to thank Guido Masarotto for discussion and advice
at various stages of preparing the manuscript. The authors are grateful
to Editor Susan Paddock, an Associate Editor, and two reviewers for
valuable comments and suggestions which greatly improved the paper.

\begin{supplement}[id=suppA]
\stitle{R Code}
\slink[doi]{10.1214/13-AOAS684SUPP} 
\sdatatype{.pdf}
\sfilename{aoas684\_supp.pdf}
\sdescription{An example of \texttt{R} code implementing beta
regression for time series analysis of Google\tsup
{\textregistered} Flu Trends.}
\end{supplement}

%

%



\printaddresses


\begin{thebibliography}{30}

\bibitem[\protect\citeauthoryear{Butler}{2013}]{butler13}
\begin{barticle}[pbm]
\bauthor{\bsnm{Butler},~\bfnm{Declan}\binits{D.}}
(\byear{2013}).
\btitle{When Google got flu wrong}.
\bjournal{Nature}
\bvolume{494}
\bpages{155--156}.
\bid{doi={10.1038/494155a}, issn={1476-4687}, pii={494155a}, pmid={23407515}}
\bptok{imsref}%
\end{barticle}
\endbibitem

\bibitem[\protect\citeauthoryear{Casarin, Dalla~Valle and
  Leisen}{2012}]{casarin12}
\begin{barticle}[mr]
\bauthor{\bsnm{Casarin},~\bfnm{Roberto}\binits{R.}},
  \bauthor{\bsnm{Dalla~Valle},~\bfnm{Luciana}\binits{L.}} \AND
  \bauthor{\bsnm{Leisen},~\bfnm{Fabrizio}\binits{F.}}
(\byear{2012}).
\btitle{Bayesian model selection for beta autoregressive processes}.
\bjournal{Bayesian Anal.}
\bvolume{7}
\bpages{385--409}.
\bid{doi={10.1214/12-BA713}, issn={1936-0975}, mr={2934956}}
\bptok{imsref}%
\end{barticle}
\endbibitem

\bibitem[\protect\citeauthoryear{Cox}{1981}]{cox81}
\begin{barticle}[mr]
\bauthor{\bsnm{Cox},~\bfnm{D.~R.}\binits{D.~R.}}
(\byear{1981}).
\btitle{Statistical analysis of time series: Some recent developments}.
\bjournal{Scand. J. Stat.}
\bvolume{8}
\bpages{93--115}.
\bid{issn={0303-6898}, mr={0623586}}
\bptnote{check related}%
\bptok{imsref}%
\end{barticle}
\endbibitem

\bibitem[\protect\citeauthoryear{Cribari-Neto and Zeileis}{2010}]{cribari10}
\begin{barticle}[auto:STB|2013/10/14|10:36:11]
\bauthor{\bsnm{Cribari-Neto},~\bfnm{F.}\binits{F.}} \AND
  \bauthor{\bsnm{Zeileis},~\bfnm{A.}\binits{A.}}
(\byear{2010}).
\btitle{Beta regression in \texttt{R}}.
\bjournal{Journal of Statistical Software}
\bvolume{34}
\bpages{1--24}.
\bptok{imsref}%
\end{barticle}
\endbibitem

\bibitem[\protect\citeauthoryear{da~Silva, Migon and Correia}{2011}]{dasilva11}
\begin{barticle}[mr]
\bauthor{\bparticle{da} \bsnm{Silva},~\bfnm{C.~Q.}\binits{C.~Q.}},
  \bauthor{\bsnm{Migon},~\bfnm{H.~S.}\binits{H.~S.}} \AND
  \bauthor{\bsnm{Correia},~\bfnm{L.~T.}\binits{L.~T.}}
(\byear{2011}).
\btitle{Dynamic {B}ayesian beta models}.
\bjournal{Comput. Statist. Data Anal.}
\bvolume{55}
\bpages{2074--2089}.
\bid{doi={10.1016/j.csda.2010.12.011}, issn={0167-9473}, mr={2785115}}
\bptok{imsref}%
\end{barticle}
\endbibitem

\bibitem[\protect\citeauthoryear{Da-Silva and Migon}{2012}]{dasilva12}
\begin{bmisc}[auto:STB|2013/10/14|10:36:11]
\bauthor{\bsnm{Da-Silva},~\bfnm{C.~Q.}\binits{C.~Q.}} \AND
  \bauthor{\bsnm{Migon},~\bfnm{H.~S.}\binits{H.~S.}}
(\byear{2012}).
\bhowpublished{Hierarchical dynamic beta model. Technical Report~253. Dept.
  Statistics, Federal Univ. Rio de Janeiro}.
\bptok{imsref}%
\end{bmisc}
\endbibitem

\bibitem[\protect\citeauthoryear{Dunn and Smyth}{1996}]{dunn96}
\begin{barticle}[auto:STB|2013/10/14|10:36:11]
\bauthor{\bsnm{Dunn},~\bfnm{P.~K.}\binits{P.~K.}} \AND
  \bauthor{\bsnm{Smyth},~\bfnm{G.~K.}\binits{G.~K.}}
(\byear{1996}).
\btitle{Randomized quantile residuals}.
\bjournal{J. Comput. Graph. Statist.}
\bvolume{5}
\bpages{236--244}.
\bptok{imsref}%
\end{barticle}
\endbibitem

\bibitem[\protect\citeauthoryear{Ferrari and Cribari-Neto}{2004}]{ferrari04}
\begin{barticle}[mr]
\bauthor{\bsnm{Ferrari},~\bfnm{Silvia L.~P.}\binits{S.~L.~P.}} \AND
  \bauthor{\bsnm{Cribari-Neto},~\bfnm{Francisco}\binits{F.}}
(\byear{2004}).
\btitle{Beta regression for modelling rates and proportions}.
\bjournal{J. Appl. Stat.}
\bvolume{31}
\bpages{799--815}.
\bid{doi={10.1080/0266476042000214501}, issn={0266-4763}, mr={2095753}}
\bptok{imsref}%
\end{barticle}
\endbibitem

\bibitem[\protect\citeauthoryear{Ginsberg et~al.}{2009}]{ginsberg09}
\begin{barticle}[pbm]
\bauthor{\bsnm{Ginsberg},~\bfnm{Jeremy}\binits{J.}},
  \bauthor{\bsnm{Mohebbi},~\bfnm{Matthew~H.}\binits{M.~H.}},
  \bauthor{\bsnm{Patel},~\bfnm{Rajan~S.}\binits{R.~S.}},
  \bauthor{\bsnm{Brammer},~\bfnm{Lynnette}\binits{L.}},
  \bauthor{\bsnm{Smolinski},~\bfnm{Mark~S.}\binits{M.~S.}} \AND
  \bauthor{\bsnm{Brilliant},~\bfnm{Larry}\binits{L.}}
(\byear{2009}).
\btitle{Detecting influenza epidemics using search engine query data}.
\bjournal{Nature}
\bvolume{457}
\bpages{1012--1014}.
\bid{doi={10.1038/nature07634}, issn={1476-4687}, pii={nature07634},
  pmid={19020500}}
\bptok{imsref}%
\end{barticle}
\endbibitem

\bibitem[\protect\citeauthoryear{Gr{\"u}n, Kosmidis and Zeileis}{2012}]{grun12}
\begin{barticle}[auto:STB|2013/10/14|10:36:11]
\bauthor{\bsnm{Gr{\"u}n},~\bfnm{B.}\binits{B.}},
  \bauthor{\bsnm{Kosmidis},~\bfnm{I.}\binits{I.}} \AND
  \bauthor{\bsnm{Zeileis},~\bfnm{A.}\binits{A.}}
(\byear{2012}).
\btitle{Extended beta regression in \texttt{R}: Shaken, stirred, mixed, and
  partitioned}.
\bjournal{Journal of Statistical Software}
\bvolume{48}
\bpages{1--25}.
\bptok{imsref}%
\end{barticle}
\endbibitem

\bibitem[\protect\citeauthoryear{Guolo and Varin}{2013}]{guolo13}
\begin{bmisc}[auto:STB|2013/10/14|10:36:11]
\bauthor{\bsnm{Guolo},~\bfnm{A.}\binits{A.}} \AND
  \bauthor{\bsnm{Varin},~\bfnm{C.}\binits{C.}}
(\byear{2013}).
\bhowpublished{Supplement to ``Beta regression for time series analysis of
  bounded data, with application to Canada Google\tsup{\textregistered} Flu
  Trends.'' DOI:\doiurl{10.1214/13-AOAS684SUPP}}.
\bptok{imsref}%
\end{bmisc}
\endbibitem

\bibitem[\protect\citeauthoryear{Hutwagner et~al.}{2003}]{hutwagner03}
\begin{barticle}[auto:STB|2013/10/14|10:36:11]
\bauthor{\bsnm{Hutwagner},~\bfnm{L.}\binits{L.}},
  \bauthor{\bsnm{Thompson},~\bfnm{W.~W.}\binits{W.~W.}},
  \bauthor{\bsnm{Seeman},~\bfnm{G.~M.}\binits{G.~M.}} \AND
  \bauthor{\bsnm{Treadwell},~\bfnm{T.}\binits{T.}}
(\byear{2003}).
\btitle{The bioterrorism preparedness and response early aberration reporting
  system (EARS)}.
\bjournal{Journal of Urban Health}
\bvolume{80}
\bpages{89--96}.
\bptok{imsref}%
\end{barticle}
\endbibitem

\bibitem[\protect\citeauthoryear{Kieschnick and
  McCullough}{2003}]{kieschnick03}
\begin{barticle}[mr]
\bauthor{\bsnm{Kieschnick},~\bfnm{Robert}\binits{R.}} \AND
  \bauthor{\bsnm{McCullough},~\bfnm{B.~D.}\binits{B.~D.}}
(\byear{2003}).
\btitle{Regression analysis of variates observed on {$(0,1)$}: Percentages,
  proportions and fractions}.
\bjournal{Stat. Model.}
\bvolume{3}
\bpages{193--213}.
\bid{doi={10.1191/1471082X03st053oa}, issn={1471-082X}, mr={2005473}}
\bptok{imsref}%
\end{barticle}
\endbibitem

\bibitem[\protect\citeauthoryear{Love et~al.}{2010}]{love10}
\begin{barticle}[auto:STB|2013/10/14|10:36:11]
\bauthor{\bsnm{Love},~\bfnm{T.~M.~T.}\binits{T.~M.~T.}},
  \bauthor{\bsnm{Thurson},~\bfnm{S.~W.}\binits{S.~W.}},
  \bauthor{\bsnm{Keefer},~\bfnm{M.~C.}\binits{M.~C.}},
  \bauthor{\bsnm{Dewhurst},~\bfnm{S.}\binits{S.}} \AND
  \bauthor{\bsnm{Lee},~\bfnm{H.~Y.}\binits{H.~Y.}}
(\byear{2010}).
\btitle{Mathematical modeling of ultradeep sequencing data reveals that acute
  CD8+ T-lymphocyte responses exert strong selective pressure in simian
  immunodeficiency virus-infected macaques but still fail to clear founder
  epitope sequences}.
\bjournal{Journal of Virology}
\bvolume{84}
\bpages{5802--5814}.
\bptok{imsref}%
\end{barticle}
\endbibitem

\bibitem[\protect\citeauthoryear{Masarotto and Varin}{2012}]{masarotto12}
\begin{barticle}[mr]
\bauthor{\bsnm{Masarotto},~\bfnm{Guido}\binits{G.}} \AND
  \bauthor{\bsnm{Varin},~\bfnm{Cristiano}\binits{C.}}
(\byear{2012}).
\btitle{Gaussian copula marginal regression}.
\bjournal{Electron. J. Stat.}
\bvolume{6}
\bpages{1517--1549}.
\bid{doi={10.1214/12-EJS721}, issn={1935-7524}, mr={2988457}}
\bptok{imsref}%
\end{barticle}
\endbibitem

\bibitem[\protect\citeauthoryear{Montgomery}{2009}]{montgomery09}
\begin{bbook}[auto:STB|2013/10/14|10:36:11]
\bauthor{\bsnm{Montgomery},~\bfnm{D.~C.}\binits{D.~C.}}
(\byear{2009}).
\btitle{Introduction to Statistical Quality Control},
\bedition{6th} ed.
\bpublisher{Wiley}, \blocation{New York}.
\bptok{imsref}%
\end{bbook}
\endbibitem

\bibitem[\protect\citeauthoryear{Ospina and Ferrari}{2012}]{ospina12}
\begin{barticle}[mr]
\bauthor{\bsnm{Ospina},~\bfnm{Raydonal}\binits{R.}} \AND
  \bauthor{\bsnm{Ferrari},~\bfnm{Silvia L.~P.}\binits{S.~L.~P.}}
(\byear{2012}).
\btitle{A general class of zero-or-one inflated beta regression models}.
\bjournal{Comput. Statist. Data Anal.}
\bvolume{56}
\bpages{1609--1623}.
\bid{doi={10.1016/j.csda.2011.10.005}, issn={0167-9473}, mr={2892364}}
\bptok{imsref}%
\end{barticle}
\endbibitem

\bibitem[\protect\citeauthoryear{Paolino}{2001}]{paolino01}
\begin{barticle}[auto:STB|2013/10/14|10:36:11]
\bauthor{\bsnm{Paolino},~\bfnm{P.}\binits{P.}}
(\byear{2001}).
\btitle{Maximum likelihood estimation of models with beta-distributed dependent
  variables}.
\bjournal{Political Analysis}
\bvolume{9}
\bpages{325--346}.
\bptok{imsref}%
\end{barticle}
\endbibitem

\bibitem[\protect\citeauthoryear{R Core Team}{2013}]{R12}
\begin{bmisc}[auto:STB|2012/06/08|12:49:54]
\borganization{R Core Team}.
(\byear{2013}).
\bhowpublished{\textit{R: A Language and Environment for Statistical
  Computing}. R Foundation for Statistical Computing, Vienna, Austria. ISBN
  3-900051-07-0. Available at \url{http://www.R-project.org/}.}
\bptok{imsref}%
\end{bmisc}
\endbibitem

\bibitem[\protect\citeauthoryear{Rocha and Cribari-Neto}{2009}]{rocha09}
\begin{barticle}[mr]
\bauthor{\bsnm{Rocha},~\bfnm{Andr{\'e}a~V.}\binits{A.~V.}} \AND
  \bauthor{\bsnm{Cribari-Neto},~\bfnm{Francisco}\binits{F.}}
(\byear{2009}).
\btitle{Beta autoregressive moving average models}.
\bjournal{TEST}
\bvolume{18}
\bpages{529--545}.
\bid{doi={10.1007/s11749-008-0112-z}, issn={1133-0686}, mr={2566415}}
\bptok{imsref}%
\end{barticle}
\endbibitem

\bibitem[\protect\citeauthoryear{Rogers et~al.}{2012}]{rogers12}
\begin{barticle}[pbm]
\bauthor{\bsnm{Rogers},~\bfnm{James~A.}\binits{J.~A.}},
  \bauthor{\bsnm{Polhamus},~\bfnm{Daniel}\binits{D.}},
  \bauthor{\bsnm{Gillespie},~\bfnm{William~R.}\binits{W.~R.}},
  \bauthor{\bsnm{Ito},~\bfnm{Kaori}\binits{K.}},
  \bauthor{\bsnm{Romero},~\bfnm{Klaus}\binits{K.}},
  \bauthor{\bsnm{Qiu},~\bfnm{Ruolun}\binits{R.}},
  \bauthor{\bsnm{Stephenson},~\bfnm{Diane}\binits{D.}},
  \bauthor{\bsnm{Gastonguay},~\bfnm{Marc~R.}\binits{M.~R.}} \AND
  \bauthor{\bsnm{Corrigan},~\bfnm{Brian}\binits{B.}}
(\byear{2012}).
\btitle{Combining patient-level and summary-level data for Alzheimer's disease
  modeling and simulation: A beta regression meta-analysis}.
\bjournal{J. Pharmacokinet. Pharmacodyn.}
\bvolume{39}
\bpages{479--498}.
\bid{doi={10.1007/s10928-012-9263-3}, issn={1573-8744}, pmid={22821139}}
\bptok{imsref}%
\end{barticle}
\endbibitem

\bibitem[\protect\citeauthoryear{Smithson and Verkuilen}{2006}]{smithson06}
\begin{barticle}[pbm]
\bauthor{\bsnm{Smithson},~\bfnm{Michael}\binits{M.}} \AND
  \bauthor{\bsnm{Verkuilen},~\bfnm{Jay}\binits{J.}}
(\byear{2006}).
\btitle{A better lemon squeezer? Maximum-likelihood regression with
  beta-distributed dependent variables}.
\bjournal{Psychol. Methods}
\bvolume{11}
\bpages{54--71}.
\bid{doi={10.1037/1082-989X.11.1.54}, issn={1082-989X}, pii={2006-03820-004},
  pmid={16594767}}
\bptok{imsref}%
\end{barticle}
\endbibitem

\bibitem[\protect\citeauthoryear{Song}{2007}]{song07}
\begin{bbook}[mr]
\bauthor{\bsnm{Song},~\bfnm{Peter X.~K.}\binits{P.~X.~K.}}
(\byear{2007}).
\btitle{Correlated Data Analysis: Modeling, Analytics, and Applications}.
\bpublisher{Springer}, \blocation{New York}.
\bid{mr={2377853}}
\bptok{imsref}%
\end{bbook}
\endbibitem

\bibitem[\protect\citeauthoryear{Stasinopoulos and Rigby}{2007}]{stasino07}
\begin{barticle}[auto:STB|2013/10/14|10:36:11]
\bauthor{\bsnm{Stasinopoulos},~\bfnm{D.~M.}\binits{D.~M.}} \AND
  \bauthor{\bsnm{Rigby},~\bfnm{R.~A.}\binits{R.~A.}}
(\byear{2007}).
\btitle{Generalized additive models for location scale and shape (gamlss) in
  \texttt{R}}.
\bjournal{Journal of Statistical Software}
\bvolume{23}
\bpages{1--46}.
\bptok{imsref}%
\end{barticle}
\endbibitem

\bibitem[\protect\citeauthoryear{Unkel et~al.}{2012}]{unkel12}
\begin{barticle}[mr]
\bauthor{\bsnm{Unkel},~\bfnm{Steffen}\binits{S.}},
  \bauthor{\bsnm{Farrington},~\bfnm{C.~Paddy}\binits{C.~P.}},
  \bauthor{\bsnm{Garthwaite},~\bfnm{Paul~H.}\binits{P.~H.}},
  \bauthor{\bsnm{Robertson},~\bfnm{Chris}\binits{C.}} \AND
  \bauthor{\bsnm{Andrews},~\bfnm{Nick}\binits{N.}}
(\byear{2012}).
\btitle{Statistical methods for the prospective detection of infectious disease
  outbreaks: A review}.
\bjournal{J. Roy. Statist. Soc. Ser. A}
\bvolume{175}
\bpages{49--82}.
\bid{doi={10.1111/j.1467-985X.2011.00714.x}, issn={0964-1998}, mr={2873791}}
\bptok{imsref}%
\end{barticle}
\endbibitem

\bibitem[\protect\citeauthoryear{Wang}{2012}]{wang12}
\begin{barticle}[mr]
\bauthor{\bsnm{Wang},~\bfnm{Xiao-Feng}\binits{X.-F.}}
(\byear{2012}).
\btitle{Joint generalized models for multidimensional outcomes: A case study of
  neuroscience data from multimodalities}.
\bjournal{Biom. J.}
\bvolume{54}
\bpages{264--280}.
\bid{doi={10.1002/bimj.201100041}, issn={0323-3847}, mr={2915307}}
\bptok{imsref}%
\end{barticle}
\endbibitem

\bibitem[\protect\citeauthoryear{Wang et~al.}{2011}]{wang11}
\begin{barticle}[mr]
\bauthor{\bsnm{Wang},~\bfnm{Weiwei}\binits{W.}},
  \bauthor{\bsnm{Scharfstein},~\bfnm{Daniel}\binits{D.}},
  \bauthor{\bsnm{Wang},~\bfnm{Chenguang}\binits{C.}},
  \bauthor{\bsnm{Daniels},~\bfnm{Michael}\binits{M.}},
  \bauthor{\bsnm{Needham},~\bfnm{Dale}\binits{D.}} \AND
  \bauthor{\bsnm{Brower},~\bfnm{Roy}\binits{R.}}
(\byear{2011}).
\btitle{Estimating the causal effect of low tidal volume ventilation on
  survival in patients with acute lung injury}.
\bjournal{J. R. Stat. Soc. Ser. C. Appl. Stat.}
\bvolume{60}
\bpages{475--496}.
\bid{doi={10.1111/j.1467-9876.2010.00757.x}, issn={0035-9254}, mr={2829186}}
\bptok{imsref}%
\end{barticle}
\endbibitem

\bibitem[\protect\citeauthoryear{Woodall}{2006}]{woodall06}
\begin{barticle}[auto:STB|2013/10/14|10:36:11]
\bauthor{\bsnm{Woodall},~\bfnm{W.}\binits{W.}}
(\byear{2006}).
\btitle{The use of control chart in health-care and public-health
  surveillance}.
\bjournal{Journal of Quality Technology}
\bvolume{38}
\bpages{89--104}.
\bptok{imsref}%
\end{barticle}
\endbibitem

\bibitem[\protect\citeauthoryear{Zou, Carlsson and Quinn}{2010}]{zou10}
\begin{barticle}[mr]
\bauthor{\bsnm{Zou},~\bfnm{Kelly~H.}\binits{K.~H.}},
  \bauthor{\bsnm{Carlsson},~\bfnm{Martin~O.}\binits{M.~O.}} \AND
  \bauthor{\bsnm{Quinn},~\bfnm{Sheila~A.}\binits{S.~A.}}
(\byear{2010}).
\btitle{Beta-mapping and beta-regression for changes of ordinal-rating
  measurements on {L}ikert scales: A comparison of the change scores among
  multiple treatment groups}.
\bjournal{Stat. Med.}
\bvolume{29}
\bpages{2486--2500}.
\bid{doi={10.1002/sim.4012}, issn={0277-6715}, mr={2897363}}
\bptok{imsref}%
\end{barticle}
\endbibitem

\end{thebibliography}
\end{document}